# Advances in infrared and imaging fibres for astronomical instrumentation

Roger Haynes[1a], Pam McNamara[b], Jackie Marcel[a], Nemanja Jovanovic[c]

[a]Anglo Australian Observatory, PO Box. 296, Epping, NSW 1710, Australia
[b]Optical Fibre Technology Centre, University of Sydney, 206 National Innovation Centre, Australian Technology Park, Eveleigh, NSW 1430, Australia
[c]Centre for Ultrahigh-bandwidth Devices for Optical Systems (CUDOS), Department of Physics, Macquarie University, North Ryde, NSW 2109, Australia

## ABSTRACT

Optical fibres have already played a huge part in ground based astronomical instrumentation, however, with the revolution in photonics currently taking place new fibre technologies and integrated optical devices are likely to have a profound impact on the way we manipulate light in the future. The Anglo Australian Observatory, along with partners at the Optical Fibre Technology Centre of the University of Sydney, is investigating some of the developing technologies as part of our Astrophotonics programme[2]. In this paper we discuss the advances that have been made with infrared transmitting fibre, both conventional and microstructured, in particular those based on fluoride glasses. Fluoride glasses have a particularly wide transparent region from the UV through to around 7µm, whereas silica fibres, commonly used in astronomy, only transmit out to about 2µm. We discuss the impact of advances in fibre manufacture that have greatly improved the optical, chemical resistance and physical properties of the fluoride fibres. We also present some encouraging initial test results for a modern imaging fibre bundle and imaging fibre taper.

## 1. INFRARED FIBRES

**1.1 Introduction**

Light transmission at any wavelength through media other than perfect vacuum incurs losses in intensity and distortion due to absorption, scattering and dispersion. Transmission without a waveguide of some sort is essentially line of sight. Light can be guided through non-linear paths by "mechanical" systems comprising arrangements of mirrors, lenses, prisms, filters and gratings or by optical fibres. Mechanical light paths are inflexible and affected by flexure, vibration, dust, moisture and temperature fluctuations. Light transmitted via optical fibres is largely protected against these effects and can be guided easily through tortuous paths and this has been used to great effect in a plethora of fibre based astronomical instruments over the past couple of decades and more. Optical fibres are not immune to flexure, mechanical stress, vibration, dust, moisture and temperature fluctuations but are typically affected much less than mechanical light paths and in different ways.

Virtually all of the fibre based instruments to date have been ground based and targeted at the visible wavelength region. However, some infrared systems have been successful. If optical fibres are to be used more extensively at infrared wavelengths, whether for ground-based or space-based instrumentation, they must be capable of transmitting efficiently over as much of the available spectrum as possible. Transparent materials with different spectral windows have been available for many years; however, some have developed a poor reputation among fibre instrument designers. Materials in current use are polymers of various types and various glasses, the principal ones being silica, fluoride and chalcogenide glasses. The materials can be made into different fibre geometries some of which have been around for many years; however, the explosion in the field of photonics fibres has introduced a seemingly bewildering choice of geometries and functionalities. These can be either solid with graduated, stepped or patterned refractive indices or microstructured with arrays of holes along the length of the fibre.

---

[1] rh@aao.gov.au; phone +61 2 9372 4815; fax +61 2 9372 4880; www.aao.gov.au
[2] This work has been funded by a PPARC Research Council Grant (UK) and was carried out in partnership with the Optical Fibre Technology Centre of the University of Sydney and University of Durham. The authors also wish to acknowledge the assistance of the Department of Physics, Macquarie University.

In the first part of this paper we hope to debunk some of the material myths and highlight some advances that have been made in infrared materials, manufacturing process and fibre structures.

## 1.2 Fibre Materials

Photonics is the science of the interactions between light and matter. The interaction can be either passive or active. Optically active fibres can modify light by amplifying it, changing its wavelength or rotating its plane of polarisation. Light can modify matter or can simply pass through and be guided from one place to another. Historically in astronomy, active modification of the incoming signal has not generally been exploited and we will only consider optically passive fibres here. However, this trend is likely to change as the need for complex and compact instrumentation develops and we consider more sophisticated fibre functionality along with integrated optical devices [1, 2].

The purpose of the passive fibre is to guide the light, protecting it from moisture and dust, with minimum loss of intensity and minimum distortion. This often requires materials which are transparent to as wide a range of wavelengths as possible and for most astronomical applications preferably with losses < 3dB/metre minimally affected by temperature change with regard to thermal expansion and refractive index[3]. The materials must be able to be drawn into fibres which are mechanically robust, flexible and unaffected by external stress or moisture. An additional requirement for possible space-based instrumentation is that the materials should be radiation hard.

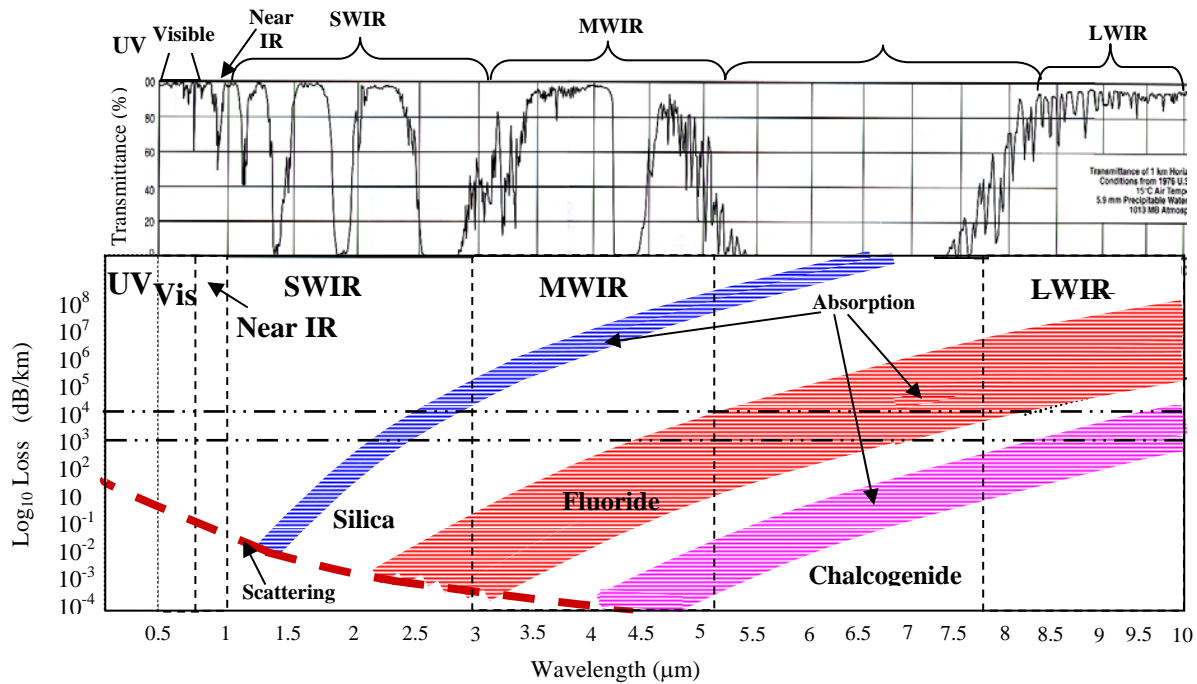

Fig. 1. The upper part of the diagram shows the atmospheric transmittance spectrum at the top of the diagram adapted from data obtained by Santa Barbara Research Center, a subsidiary of Hughes. The lower part of the diagram shows the average ranges of optical losses in silica, fluoride and chalcogenide glass fibres compared with the spectral transmittance of the atmosphere. In theory silica can be used up to ~2.0 - 2.5μm, fluoride glasses up to ~4.5 - 7μm, depending on composition and chalcogenide glasses up to ~8 - 10 μm, depending on composition. Intrinsic scattering loss that dominates the UV is roughly the same for all three-glass systems.

---

[3] Low optical dispersion is not typically considered a pre-requisite for astronomical instrumentation, however, the wavelength dependent phase shift dispersion introduces may be an issue for some applications such as interferometry.

A schematic comparison between the theoretical transmission losses of three types of glasses – silica, fluoride and chalcogenide is shown in Fig. 1. Note that actual losses in real glasses are almost invariably higher than those represented here. Optical loss is a combination of scattering and absorption due to both intrinsic properties of the fibre materials and extrinsic effects caused by defects and microstructures and nanostructures in the fibre geometries. The degree of scattering is dependent on the size of the scattering centre relative to the wavelength of light transmitted and Rayleigh scattering from molecular scale structures predominates at shorter wavelengths and is much the same for all materials. Rayleigh scattering is proportional to $d^6/\lambda^4$, where d is molecule size and $\lambda$ the scattered wavelength. Reduction of mean molecular size in the fibre material is the only way to reduce loss in the ultraviolet and some tuning of the structures is possible through both the manufacturing process and the choice of the molecular composition of the glass. Absorption takes over at longer wavelengths and is different for different materials. In real glasses scattering and absorption losses due to defects, impurities and geometric effects are added to the fundamental losses inherent in the materials.

### 1.3 Silicate glasses

Most optical fibres on current use worldwide are based on silica glass and this is also the case in astronomy. Fused silica ($SiO_2$ glass), basically a single-compound glass, is mechanically robust and unlike most glasses highly resistant to thermal shock. It is the most stable[4] of all possible glasses and is transparent to wavelengths from ~0.3 to 2.5μm. The optical properties of silicate glasses have a limited range, since only low levels of a few other elements can be added to it. Easy modifiers include germanium, phosphorus and boron; anything else is difficult. Silica fibre has been developed for telecommunications for about 40 years since its use for this purpose was first suggested in 1966 [3]. It was not until the early 1980s that silica fibre was considered sufficiently well-developed to displace copper in telecommunications networks. Silica fibre can now be made with optical transmission losses as low as 0.02dB/km, close to the intrinsic limit of silica, though not all fibre produced has such low loss.

It is commonly believed that silica fibre lasts for ever, but in practice its mechanical strength degrades quite rapidly due to moisture penetration into surface micro-cracks and it is prone to radiation-damage by α- and β-particles and high power laser light (Table 1). This degradation may not only involve the mechanical properties of the fibre, it may also affect its optical properties.

### 1.4 Fluoride glasses

Fluoride glasses are multi-component glasses based entirely on fluoride compounds, containing no oxides or other anionic species. Heavy metal fluoride glasses (HMFG) were discovered accidentally in the early 1970s [4]. This family of glasses are intrinsically less robust than silica, and can be more prone to extreme thermally shock and crystallisation. Theoretically they promised much higher transparency than silica (~0.001dB/km) and for many years telecommunications companies pursued them for that reason alone. However, practical achievement of such low losses proved difficult so they were largely abandoned by telecommunications in the 1990s. Some researchers persisted and high quality fluoride glasses and fibres are now available. In general, fluoride glass fibres have the requisite transparency for astronomy from the ultraviolet to beyond 6μm and have a very wide range of properties which can be tailored to fit particular applications by adjusting glass composition [5]. There are three principal groups of fluoride glasses – the fluorozirconates, based on $ZrF_4$, the fluoroaluminates, based on $AlF_3$ and the fluorogallates, based on $GaF_3$. The fluorozirconate ZBLAN glasses ($ZrF_4$:$BaF_2$:$LaF_3$:$AlF_3$:NaF) are the best-known, but suffer from susceptibility to attack by water, mainly due to inclusion of water-soluble NaF in their composition. It was principally this proneness to water corrosion that gave fluoride glasses a bad reputation. This problem can be overcome by reducing the water soluble components in the glasses and ensuring exclusion of water from the glass fabrication process and subsequent environment Fluoride glasses and fibres can be used for all wavelengths between ultraviolet and out as far as ~7μm for some fluoride glasses. Fluoride glasses have the additional advantage of being significantly more resistant to radiation damage (Table 1) making them more suitable for space-based applications. Mechanical strength of fluoride fibres is initially about half that of silica when both are newly drawn but degrades more slowly, so that after about 5 years exposure the two types of fibre are equal and thereafter silica fibres may become the weaker of the two. Figure 3 shows a fluoride glass (ZBLAN) fibre twisted tightly around a matchstick in a simple demonstration of its bend strength. Hoya Corporation of Japan has developed a glass to carry high laser power for dental surgery [6] which is a combination of

---

[4] Stable in this case means least likely to become crystalline. This would introduce scattering from the grain boundaries.

AlF$_3$, ZrF$_4$, NaCl and a number of other fluorides. Despite the high solubility of NaCl the overall composition is less sensitive to water than standard ZBLAN.

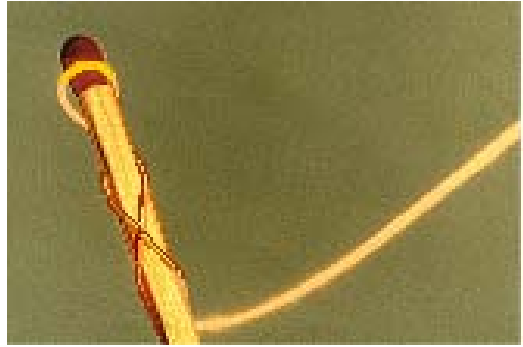

Courtesy D.Coulson

Fig. 2. ZBLAN fluoride glass fibred twisted round a matchstick.

**1.5 Chalcogenide glasses**

Chalcogenide glasses have been explored since the 1960s and are based on compounds of the Group VI elements S, Se and Te, usually in combination with Group IV and V elements. Oxygen is also a chalcogenide but oxide glasses are treated as a separate category. Chalcogenide glasses theoretically have lower optical losses than fluoride glasses beyond ~3μm and could extend out to ~10μm. In practice chalcogenide glasses can be difficult to fabricate with these low losses, though great improvements have been effected in recent years [7, 8]. The chalcogenides would be useful in space-based applications as some forms are exceeding radiation hard (Table 1). Researchers at the Naval Research Laboratories (NRL) in the USA have produced chalcogenide glass fibre which can transmit very high optical power without laser damage (Table 1). This fibre is now marketed by CorActive Specialty Optical Fiber Manufacturers (USA). On CorActive's Webpage [9] it is said to have a broad range of core diameters from 8 to 600 μm and power capacity of 1 GWcm$^{-2}$ for pulsed and 1 MWcm$^{-2}$ for continuous wavelengths. CorActive claim that their infrared fibres transmit wavelengths from ~ 1.5 to 9μm with losses less than 0.6dBm$^{-1}$, other than an attenuation spike (~ 8dBm$^{-1}$) at around 4.8μm. In general chalcogenide glasses are not prone to water attack but can be toxic when they contain arsenic or antimony. Unprotected chalcogenide surfaces can degrade quite readily. Its initial mechanical strength is approximately the same as silica.

The mechanical strength and radiation hardness in chalcogenide and fluoride glasses are probably dependent more on quality control during fabrication rather than glass composition. Some of the early fluoride fibres were particularly poor in this regard, however, many manufactures have made huge improvements in their processes in recent years.

Table 1. Laser Damage Thresholds in Glasses [9 - 13]

| Glass Type | Max. Power Density (W/cm$^2$) |
|---|---|
| NRL chalcogenide | 477 x 10$^6$ (>1.1 x 10$^9$) |
| Hoya AZYRN, AlF$_3$:ZrF$_4$-chlorofluoride | 8 x 10$^6$ |
| As-S chalcogenide | 0.2 x 10$^6$ |
| Silica | 0.1 x 10$^6$ |
| Tm:Ho-doped ZBLAN | 0.042 x 10$^6$ |
| GeSeTe chalcogenide | 0.04 x 10$^6$ |

**1.6 Polymers**

Highly flexible fibres can be made from transparent polymers such as PMMA (polymethylmethacrylate) but these tend to be mechanically weaker than glass fibres. Their bending properties are typically better than glasses, but they lack the tensile strength of glass. However, for many applications the flexibility of these fibres can be an advantage. Polymers are also useful materials for prototyping and experimenting with novel fibre structures as they can be much easier to machine and handle during the fibre preform stages. Their spectral window is approximately the same as silica, between ultraviolet and 2μm and their minimum optical loss is in the region of 1dB/metre for solid core fibres and this is close to the intrinsic limit for such materials.

**1.7 Hollow Fibres**

None of the above materials is transparent beyond ~11μm. The only possibility at present is hollow fibre [14], effectively a tube with a reflective inner surface, designed to force transmitted light to travel in the large air-filled core. Its ability to transmit very long wavelength depends on the transparency of clean dry air to these. Vacuum-cored hollow fibres might be better and could be a natural consequence of using such fibres in space based applications. Such fibres are available and either relies on grazing incidence or reflection from coatings on the inside of the hollow tube. A variant of this idea will be discussed below with regard to air-cored microstructured fibres.

**1.8 Fibre geometries**

The aim of most passive optical fibre is to ensure that light coupled into the fibre does not escape even when the fibre is bent. This is achieved by various microstructures which confine the transmitted light in the fibre core. The simplest geometry is the step-index fibre, in which a cylindrical core of high refractive index material is clad in a sheath of lower refractive index; historically this type of fibre has been used for spectroscopic applications in astronomy. Light launched into the core experiences total internal reflection at the interface between core and cladding and is prevented from passing out through the wall of the fibre. In other fibres the refractive index is graduated, decreasing with distance from the fibre centre (Graded index fibres). A different kind of fibre has been developed in the last ten years which uses an array of holes, traversing the entire length of the fibre. These are known variously as "holey fibres", microstructured optical fibres (MOFS) or photonic crystal fibres (PCFs). The holes can be thought of as a way of mixing low refractive index air with higher refractive index glass in the cladding, thus creating a region of effective lower refractive index around the core. An even more interesting development is the "air-cored" microstructured optical fibres (AMOFs), sometimes known as Photonic Band Gap (PBG) fibres. The hole array in these acts to reflect the transmitted light back into the air-core at the centre, even though this has lower refractive index than the surrounding cladding. The PBG fibres typically limit wavelength regions in which they propagate light efficiently. A few of the possible hole array patterns are shown in Fig. 3.

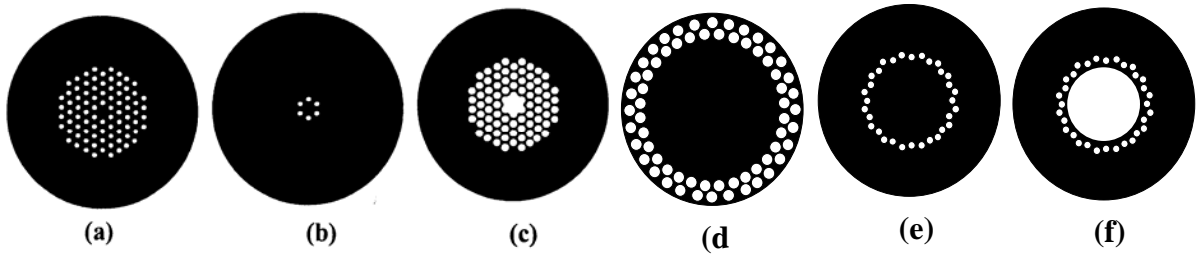

Fig. 3. Fibre cross-sections - some microstructured optical fibre hole arrays (aka "holey fibres" or PCFs – Photonic Crystal Fibres). (a), (b) (d) and (e) are solid core fibres, (c), and (f) are "air-cored" or PBG – Photonic Band Gap fibres.

Microstructured fibres can be designed to be single-moded for almost all wavelengths even when the effective diameter of the core is very large. To be more precise, the fibre is single-moded because all modes except one leak out of the fibre through the gaps between the holes. This enables very high single-mode power densities to be carried in fibre materials which do not suffer laser damage. Microstructured fibre design can also be multi-moded and among many attributes possibly useful to astronomy the structures can impart to the fibres is a very high numerical aperture.

Air-cored microstructured fibres effectively transmit light through air, which when dry and dust-free has much lower attenuation than any glass or polymer. AMOFs are sometimes referred to as material agnostic – meaning that their properties are independent of the material of which the solid part of the fibre is made. This is not quite true, but a good approximation. The holes in optical fibres can also be filled with other media – gases other than air, liquids, even solids of different refractive indices. It might even be possible to evacuate the holes – and a vacuum attenuates light even less than air. PBG AMOFs are very difficult to fabricate because their properties depend critically on the precise geometry of the hole arrays. A successful fibre of this type, similar to the design shown in Fig. 3(f), has been made in polymer [15] but this has not yet been transferred to glass. If the effect is truly "material-agnostic" this may not matter and this mPOF (microstructured Polymer Optic Fibre) fibre could be used at wavelengths beyond 11µm and the capabilities of the available glasses. This is, in effect, a variant of the hollow fibre mentioned above in sub-section 2.1.5. The rings of holes around the large central hole act as cylindrical Bragg gratings and reflect light back into the hollow core.

### 1.9 Astronomical applications

In astronomical spectroscopy, multi-mode fibres have revolutionised multi-object survey instruments with systems such as 2dF [16] and SLOAN [17]. Next generation multi-object systems such as the proposed WFMOS [18] instrument, with plans for many thousands of fibres, continue to exploit the "light pipe" properties of fibres. Fibres have also been used for both large IFUs such as the VIMOS IFU [19], with ~6800 fibres, and smaller deployable IFUs as employed by the VLT-FLAMES system. Virtually all of the spectroscopic applications have been limited to the visible regime, with only a few applications (such as the FMOS-Echidna [20] system, soon to be commissioned on the Subaru telescope) working in the infrared. However, the infrared applications typically operate at wavelengths short-wards of 2µm. Optical fibres have been shown to remain mechanically flexible and robust at cold temperatures [21] typically required for systems working beyond 2µm. With careful system design such cold environments should not significantly affect their optical properties either. With the recent development of infrared materials, the fibre manufacturing process and fibre geometries there should be increased opportunities to extend the wavelength of fibre based instruments well beyond 2µm out to 11µm and beyond, potentially revolutionising fibre based instruments for both ground based and space based systems.

Single mode fibres have typically been used in astronomy in a similar ways to the telecoms industry, i.e. for data transport and communications. However, single mode fibres are now being used as spatial filters for interferometry to remove the impact of random phase shifts introduced by atmospheric turbulence, dramatically improving coherence measurement. This technique was pioneered by the French FLUOR instrument [22]. The development of endless single mode MOFS with selectable effective core sizes could both simplify and improve coupling efficiency into the fibres. The infrared material and endless single mode developments might be exploited to extend the wavelength over which such systems operate efficiently.

For interferometry the frequently poor efficiencies with which the telescope can be coupled to the single mode fibre is more than compensated for by the gains in coherence measurement. For spectroscopy, the sort of coupling losses (up to ~99 % or more for a poorly matched system) that result from the impact of atmospheric turbulence on the incoming wavefront [23 & 24] and matching the fibre to the telescope optics are typically unacceptable. However, the amount of light typically "thrown away" by coupling a ground based telescope into a single mode fibre can be reduced dramatically by improving or correcting the atmospheric turbulence using adaptive optics (AO) systems and matching the fibre core parameters to suit the telescope focal station. With the promise of extreme AO and multi-conjugate AO systems on large telescopes delivering near diffraction limited performance, the coupling efficiency into single mode fibres should improve sufficiently to consider them practical for non interferometric applications. This opens up a whole "raft" of possibilities that might exploit some of the multitude of functionality available (and promised) by active fibres and integrated optical devices. One exciting new development for astronomical imaging and spectroscopy that can be greatly simplified by the developments in AO systems are the OH Suppression fibres [1] being developed at the Anglo-Australian Observatory. These fibres, that block out the night sky lines, can work in as multi-mode fibres, but can be greatly simplified if only required to operate as single mode fibres.

Much of field of active fibres technologies and integrated optics, that will revolutionise instrumentation [2] in the next few years, is targeted at single moded applications with much of the functionality being extremely challenging or impractical to implement in a multi-mode regime.

## 2. IMAGING FIBRE BUNDLES

### 2.1 Introduction

Imaging fibre bundles (IFBs) (in astronomy sometimes called coherent fibre bundles) have been around for many years, but have been subject to continued development and we now have available bundles with many thousand of elements. The development has been driven by the requirement for high spatial resolution imaging capabilities in flexible, hermetic and convenient optical trains, and the performance of optical transformation or re-formatting, in robust and convenient to use devices. Some applications for these imaging bundles include endoscopy for the biomedical industry, remote monitoring for surveillance and image monitoring in hostile environments.

Our original goal was to define the imaging properties of these bundles and to see if they are suitable for scientific applications. The initial results look promising, but significant further work needs to be done.

## 2.2 Image fidelity

IFBs are available from companies such as Sumitomo Electric [25] and Schott. An example of the structure of the active region of a Sumitomo Electric imaging fibre bundle is shown in Fig. 4 at various magnifications.

Sumitomo Electric specifies that the fibre elements typically have numerical apertures (NAs) ranging from 0.3 to 0.35 and fibre densities on the order of $10^4$/mm$^2$. The bundles themselves are structurally flexible (though during handling we noted that the 20/50 IFB has significant rigidity), with allowable bend radii of 10-100mm, increasing with the bundle core size. The initial image fidelity test pictures as shown in Fig. 5. The images where obtained with the Sumitomo Electric IFBs 20/50 fibre (with approximately a 1.8mm diameter active region). Test targets were back illuminated and placed directly in front of the input end face of the bundle in order to gauge how well the IFB reproduced the target image. We were able to resolve structure at around the 5μm level (Fig. 5 left).

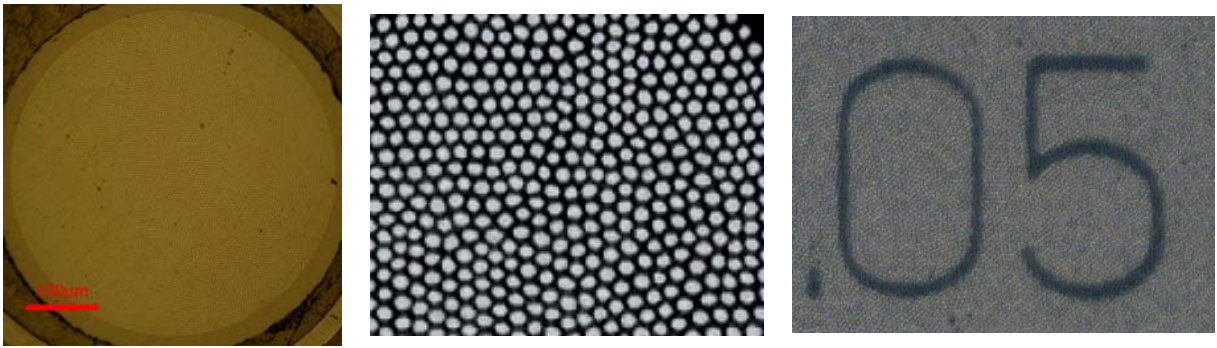

Fig. 4. Cross-sectional view of a 500μm diameter Sumitomo Electric IFB with 10,000 elements (left). A highly magnified back illuminated image of the Sumitomo Electric IFB end-face showing the semi-hexagonally packed structure of the fibre elements within the core (middle). Magnified image of part of a microscope graticule imaged through a Sumitomo Electric IFB, the pixilation due the discrete fibre cores (right).

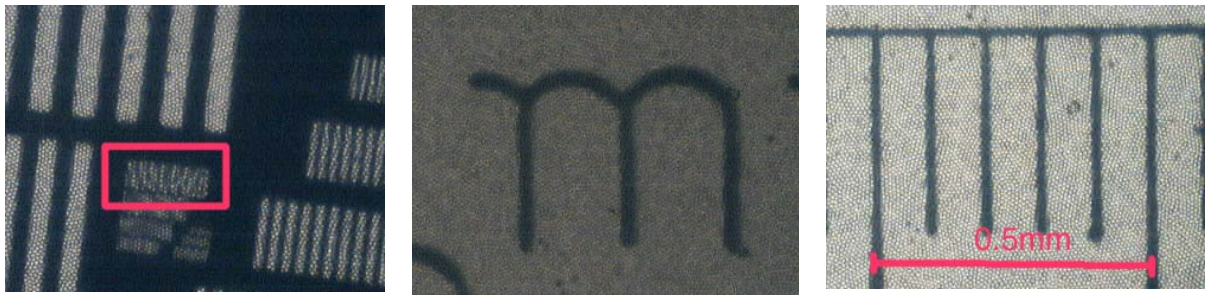

Fig. 5. Images through a Sumitomo Electric 20/50 image guide bundle of a high-resolution target shows resolved detail down to ~10μm (left, highlighted area). Examples of images of a graticule through the same 1.8mm core, 50,000 fibre element image guide bundle (middle and right).

Schott produces both IFB and imaging fibre tapers. We obtained an imaging fibre taper with aspect ratio of 1:4 that can either magnify/de-magnify the image. These fibre tapers are used for a number of advanced imaging applications including; CCD and image intensifier coupling, medical and dental radiography, video imaging, fluoroscopy and more. The initial image fidelity test pictures are shown in Fig. 6. The images of a high resolution test target taken through the Schott imaging taper and appear of high quality. They do not show the graininess apparent in the Sumitomo Electric IFB test pictures (Fig. 5).

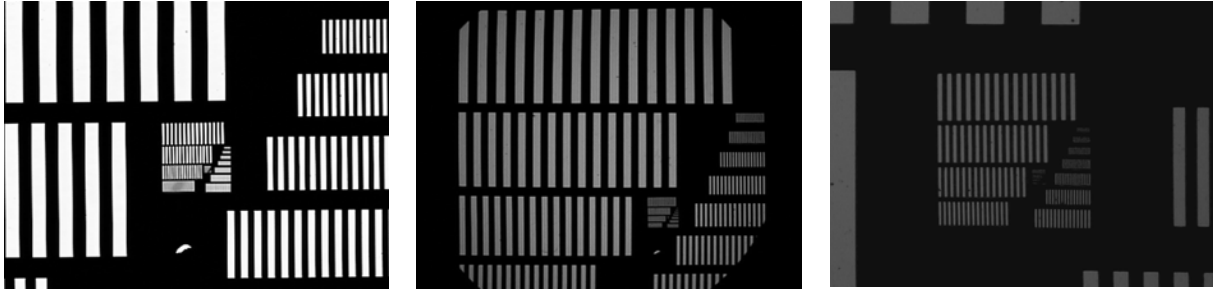

Fig. 6. Images of a high resolution test target taken using a Micro-Vu microscope/measuring machine. Left – Test target imaged using microscope. Middle – Test target de-magnified (4:1) through a Schott imaging fibre taper. Right – Test target magnified (1:4) through a Schott imaging fibre taper.

### 2.3 Cross talk and scattering

For some possible applications in astronomy the IFBs photometric integrity would be important. We therefore set up a test in which a back-illuminated 100μm pinhole was place at the input end face of the fibre and image recorded at the IFB output face. The pinhole was illuminated separately with a green LED (λ ~ 525nm) and then a red LED (λ~ 625nm). The initial radial profiles of the output images for the 1.8mm diameter Sumitomo Electric IFB (20/50) look promising. The radial profile, shown in Fig. 7, has been mirrored for illustration purposes. The significant structure present towards the centre of the profile is a result of the pixilation at the output introduced by the fibre elements. The structure is reduced (as square root of the radius) by the averaging effect of integrating over many pixels as the radius increases. A key issue here is the element to element uniformity. Often that mottling would be input at the detector in astronomical observations. This is reminiscent of imaging through birefringent materials [27] like $LiNO_3$ (e.g. tunable solid etalons), in that case placing the etalons in the illumination pupil can side step the problem. Other techniques to reduce the structural impact of the fibre elements, which can be used at the image plane, are dithering and binning. With most of the power contained within 100μm the level of cross talk to adjacent regions of the IFB appears to be small and encouraging enough to warrant further investigation of the IFB characteristics for astronomical applications.

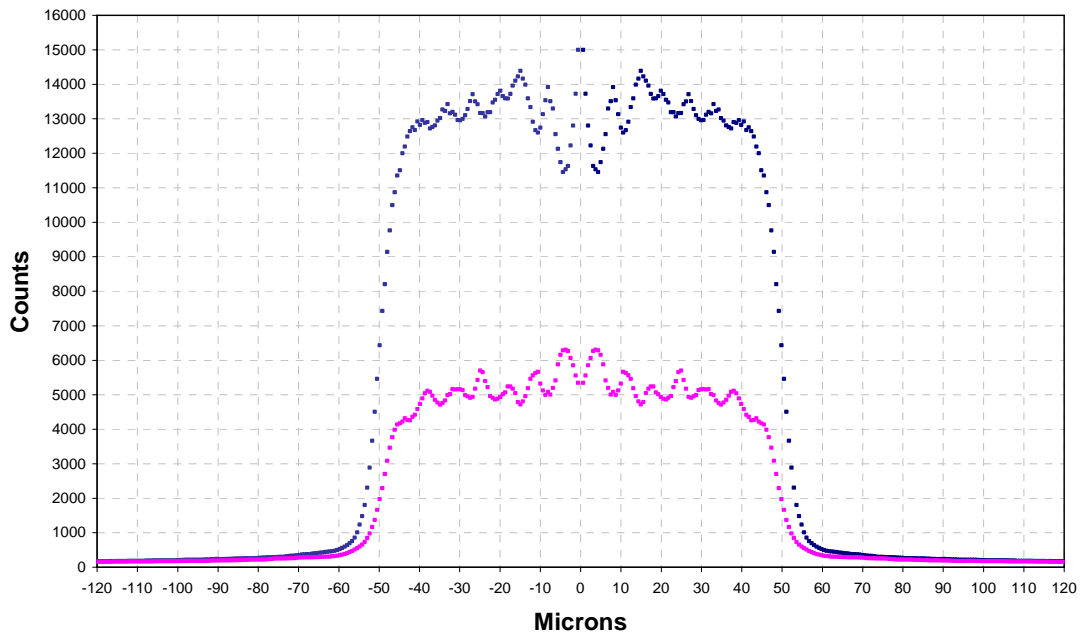

Fig. 7. Mirrored radial profile of the output image from a Sumitomo Electric IGN-20/50 imaging fibre bundle (1.8mm core diameter, ~2.0m long) with a 100μm diameter spot projected onto the input face. The upper line corresponds to the green LED and the lower line the red LED (the green LED was intrinsically brighter) The average detector counts in the annulus (y-axis) at a given radius (x-axis) from centroid are given for the image at the output face of the fibre bundle.

**2.4 Attenuation**

The Sumitomo Electric IFB claim attenuation of <0.35dB/m or better over the visible wavelength region. However, such measurements are typically determined using a cut back method (commonly use in the telecommunications sector), this only accounts for losses in the bulk material and does not typically include coupling losses that can occur through poor modal coupling, filling factor effects and reflection losses. System throughput is often a critical parameter for astronomical instrument and given the encouraging imaging performance of these devices we now plan to investigate the Schott taper and the Sumitomo Electric IFBs full transmission characteristics as a function of wavelength.

**2.5 Possible astronomical applications**

The challenges presented in wide field multi-object astronomy (imaging and spectroscopic) have long required instrument designers to develop novel ways to sample the potentially overwhelming number of spatial elements within the required field of view (FOV). As telescope systems have become more sophisticated, with the advent of adaptive optics, the sort of field and imaging performance promised with extremely large telescopes (ELTs) it can be totally impractical [26] (both technically and financially) to sample every element (potential up to $10^9$ spatial elements in the available FOV), forcing astronomer to selectively sample those regions of specific interest within the available FOV. This can potentially reduce the opportunities for serendipitous discoveries, however, for many science programmes this is not a concern and so there is a multitude of instruments and designs that exploit the discrete sampling technique over a wide range of observing regimes, such as 2dF [16], VIMOS [28], KMOS [29], WFMOS [18], Honeycomb [30]. There are other cases where instrument design drivers make it highly desirable for the main instrument to be mounted remotely from the telescope, such as the high spectral resolution systems bHROS [31] and HARPS [32]. With ELT concepts there could be additional constraints imposed by the location of the focal station, the mass and size constraints as well as the instrument requirements. For example, the now defunct OWL ELT design had the location of focal stations within the optical train of the telescope, providing strong impetus to move instrument off-telescope.

A good example to demonstrate possible applications for imaging bundles might be the Starbug concept [33]. The AAO concept development (previously part of the Opticon - Smart Focal Plane [34] development) is a generic focal surface positioning system in which an arbitrary number of Starbug units can be randomly deployed over the focal surface in order to sample sub-areas within the total FOV. These Starbug units could perform a variety of functions from simple optical pick-offs to a fully integrated instrument platform [35]. However, one could imagine how a simple device that could relay the focal plane image with little loss of throughput and image fidelity to a remote location away from the focal surface, and furthermore possibly magnify/de-magnify and/or transform/re-formate the image (in an integrated device without free space optics). This could be highly advantageous. Some, and in future all, of these functions may be possible with imaging fibre bundles, imaging taper and the like. Some of the many potential advantages of such technologies are tolerant of movement and flexure of the intervening bundle (the image position at the output depends on the image position at the input) and design flexibility in controlling the location of the image and pupil plane within the instrument optical train. These could all allow extremely simple optical trains, of many metres in length, to be easily implemented in robust, hermetically-sealed, clean and relatively stable fibre bundles. Furthermore, image scattering, ghosting and stray light issues are greatly simplified and may be easier to control.

Even to provide a simple alternative to the traditional "home-made" fibre bundle for multi-fibre, systems consisting of only a few fibres (typically a minimum of 7 for guide fibres, e.g. the 2df, OzPoz [36] guide fibres) could be valuable. For fibre based IFUs (such as VIMOS with ~6800 fibres) one could envisage more effectively purchasing these sort of devices "off the shelf", containing possible tens of thousands of fibre elements.

## 3. SUMMARY

Infrared fibre materials and manufacturing have made significant improvements over the past decade. These have enhanced the wavelength range of infrared fibre material from visible wavelengths through to 11µm and beyond. The industry has also addressed the historic issues and concerns with regards the robustness and handling of these infrared material. The material and manufacturing techniques are now in place such that robust and broad wavelength coverage fibres can be manufactured. These developments, along with the vast parameter space opened up by new fibre geometries, could pave the way for a revolution in infrared instrumentation, giving opportunities for science programmes that might rival the impact optical fibres have had in visible wavelength astronomy. With the radiation hardness of these materials also improving they could be suitable for both ground-based and space-based astronomy projects.

There have also been great advances in imaging fibre bundles. The AAO has carried out initial image fidelity tests on devices from two manufactures and these have been very encouraging, warranting investment in further testing in order to assess the suitability of such devices for scientific applications.

# REFERENCES

1. J. Bland-Hawthorn, M. Englund, and G. Edvell, "New approach to atmospheric OH suppression using an aperiodic fibre Bragg grating", *Opt. Express 12*, 5902-5909 (2004).

2. J. Bland-Hawthorn & A. Horton, "Instruments without optics: an integrated photonic spectrograph", *Ground-based and Airborne Instrumentation for Astronomy, ed. I. McLean & M. Iye*, Proc. SPIE **6269**, paper 23, 2006.

3. K.C. Kao and G.A. Hockham, "Dielectric-fibre surface waveguides for optical frequencies" *Proc. I.E.E.*, 113, 1151-1158 (1966).

4. M. Poulain, M. Poulain and J. Lucas, :Fluorine-containing glass with $ZrF_3$: Optical properties of a glass doped with $Nd^{3+}$ ", *Materials Research Bulletin,* 10(4), 243-6 (1975).

5. P. McNamara and R.H. Mair, "Devitrification Theory and Glass-Forming Phase Diagrams of Fluoride Compositions", *Proc. of the SPIE, Smart Materials, Nano- and Micro-Smart Systems* (12-15 Dec 2004), 5650 123 -34 (8 March 2005).

6. T. Yamashita, "Recent advances in IR-transmitting fibers for laser power delivery", *Review of Laser Engineering*, 27(3) 167-72 (1999).

7. D. Lezal, J. Pedlikova and J. Zavadil, "Chalcogenide glasses for optical and photonics applications", *Chalcogenide Letters,* 1(1), 11 – 15 (2004).

8. T.M. Monro, Y.D. West, D.W. Hewak, N.G.R. Broderick and D.J. Richardson, "Chalcogenide holey fibres", *Electronics Lett.* 36(24) 1998 – 2000 (2000).

9. F. Chenard, "Advanced Mid-IR and DCOF", *http://www.coractive.com*

10. E.M. Dianov, I.A. Bufetov, A.A. Frolov, V.V. Plotnichenko, A.V. Shubin, M.F. Churbanov and G.E. Snopatin, "Catastrophic damage in special fibers", *OFC2002, paper 323*.

11. M. Doshida, K. Teraguchi and M. Obara, "Gain measurement and upconversion analysis in $Tm^{3+}$, $Ho^{3+}$ co-doped alumino-zirco-fluoride glass", *IEEE Journal of Quantum Electronics* 131(5) 910 – 15 (1995).

12. Yan Feng, Xiaobo Chen, Feng Song, Kun Li and Guanyin Zhang: "Upconversion luminescence of ZBLAN:$Tm^{3+}$,$Yb^{3+}$ glass pumped by a ~970 nm LD and its concentration effect", *Proc. of the SPIE* 3551 116 – 20 (1998).

13. E.M. Dianov, V.G.Plotnichenko, Yu.N. Pyrkov, I.V. Smol'nikov, S.A. Koleskin, G.G. Devyatykh, M.F. Churbanov, G.E. Snopatin, I.V. Skripachev and R.M. Shaposhnikov, "Single-mode As-S glass fibers", *Inorganic Materials* 39(6) 627-30 (2003). Pub. MAIK Nauka, Russia.

14. J.A. Harrington, "The development of hollow core waveguides for in telecommunications, sensors, and laser power delivery", *Optical Fiber Communication Conference (OFC) (IEEE Cat. No.04CH37532) Washington, DC, USA, Opt.Soc. America.* 3 (2004).

15. A. Argyros, M.A. van Eijkelenborg, M.C.J. Large, I.M. Bassett, "Hollow core microstructured polymer optical fibres," Optics Letters 31(2), 172-4 (2006).

16. I.Lewis et al., "The Anglo-Australian Observatory's 2dF facility", *Mon. Not. R. Astron. Soc.* **333 issue 2**, 279-299, 2002

17. D. G. York, et al., "The Sloan Digital Sky Survey: Technical Summary", *The Astronomical Journal,* Volume 120, Issue 3, 1579-1587, 2000.

18. S.Barden et al., "WFMOS: a feasibility study for Gemini", *Ground-based and Airborne Instrumentation for Astronomy, ed. I.McLean & M.Iye*, Proc. SPIE **6269**, paper 76, 2006

19. E. Prieto, O. Le Fevre, M. Saisse, C. Voet, C. Bonneville, "Very wide integral field unit of VIRMOS for the VLT: design and performances", *Optical and IR Telescope Instrumentation and Detectors, ed. M.Iye & A.Moorwood*, Proc. SPIE **4008**, 510-521, 2000.

20. P.Gillingham et al., "Echidna – A multi-fiber positioner for the Subaru prime focus", *Optical and IR Telescope Instrumentation and Detectors, ed. M.Iye & A.Moorwood*, Proc. SPIE **4008**, 1395-1403, 2000

21. Lee, D., **Haynes, R.,** Skeen, D.J., "Properties of optical fibres at cryogenic temperatures", *MNRAS*, Vol 326, pp774-780, 2001.

22. Coude du Foresto V. et al. "FLUOR fibered instrument at the IOTA interferometer", *Astronomical Interferometry, ed. R. D. Reasenberg,* Proc. SPIE **3350**, 856-863, 1997.

23. S. Shaklan, F. Roddier, "Coupling starlight into single-mode fiber optics", *Applied Optics,* Vol. 27, No. 11, 2334-2338, 1988.

24. V. Coude du Foresto, M. Faucherre, N. Hubin, P. Gitton, "Using single-mode fibers to monitor fast Strehl ratio fluctuations", *Astronomy & Astrophysics Supplement Series,* **145,** 305-310, 2000.

25. Sumitomo Electric, "Image Guide Specialty Fiber Products", http://www.sei.co.jp/optproducts/products/fiber/imageg.html

26. A.Russell et al., "Instruments for a European Extremely Large Telescope: The challenges of designing instruments for 30-100m telescopes", *Ground-based Instrumentation for Astronomy, ed. A.Morwood & M.Iye*, Proc. SPIE **5492**, 1796-1809, 2004.

*27.* J. Bland-Hawthorn, "Tunable Imaging Filters", *Encyclopaedia of Astronomy and Astrophysics ed. Paul Murdin,* article 5401, Bristol: Institute of Physics Publishing, 2001.

*28.* O. Le Fevre et al. "Commissioning and performance of the VLT-VIMOS", *Instrument Design and Performance for Optical/Infrared Ground bases Telescope, ed. M. Iye, A. Moorwood*, Proc. SPIE **4841**, 1670-1681, 2003.

29. R.Sharples et al., "KMOS: an infrared multiple-object integral field spectrograph for the ESO VLT", *Ground-based Instrumentation for Astronomy, ed. A.Morwood & M.Iye*, Proc. SPIE **5492**, 1179-1186, 2004.

30. J.Bland-Hawthorn et al., "Honeycomb: a concept for a programmable integral field spectrograph", *Ground-based Instrumentation for Astronomy, ed. A.Morwood & M.Iye*, Proc. SPIE **5492**, 242-250, 2004.

31. M. E. Aderin, "bHROS installation and system performance", *Ground-based Instrumentation for Astronomy, ed. A.Morwood & M.Iye*, Proc. SPIE **5492**, 160-171, 2004.

32. HARPS Project, http://www.ls.eso.org/lasilla/sciops/3p6/harps/

33. R.Haynes et al., "It's alive! Performance and control of prototype Starbug actuators", *Optomechanical Technologies for Astronomy,* ed. E.Atad-Ettedgui, J.Antebi & D.Lemke, Proc. SPIE **6273**, paper 69, 2006.

34. Colin Cunningham, "Smart Focal Planes" http://www.astro-opticon.org/presentations/misc_downloads/JRA5_18_month_report.pdf

35. A.McGrath and R.Haynes, "Deployable payloads for Starbug", *Optomechanical Technologies for Astronomy, ed. E.Atad-Ettedgui, J.Antebi & D.Lemke*, Proc. SPIE **6273**, paper 70, 2006.

36. P.Gillingham et al., "The performance of OzPoz, a multi-fiber positioner on the VLT", *Instrument Design and Performance for Optica/Infrared Ground-based Telescopes, ed. M.Iye & A.Moorwood*, Proc. SPIE **4841**, 1170-1179, 2003